\documentclass[preprint,showpacs]{revtex4}
\usepackage{graphicx}%
\usepackage{dcolumn}
\usepackage{amsmath}
\usepackage{latexsym}

\begin{document}
\title{Atom Interferometry and the Gravitational Redshift   }  
\author{Supurna Sinha and Joseph Samuel}
\affiliation{Raman Research Institute, Bangalore-560 080}
\date{\today}
\begin{abstract}

From the principle of equivalence, Einstein
predicted that clocks slow down in a gravitational field. Since
the general theory of
relativity is based on the principle of equivalence, it is   
essential to test this prediction accurately.
M\"uller, Peters and Chu claim that a reinterpretation of
decade old experiments with atom interferometers leads to a sensitive
test of this gravitational redshift effect at the Compton frequency.
Wolf {\it et al} dispute this claim and adduce arguments against it.
In this article, we distill these arguments to a single fundamental
objection: an atom is {\it not} a clock ticking at the Compton
frequency.
We conclude that atom interferometry experiments conducted to date do 
not yield such sensitive tests of the gravitational redshift.
Finally, we suggest a new interferometric experiment to measure the
gravitational redshift, which realises a quantum version of the 
classical clock ``paradox''.

\end{abstract}

\pacs{04.80.-y,03.75.Dg}
\maketitle



\section{Introduction}
The gravitational redshift is the
first prediction \cite{Schwartz} made by Einstein from the Principle of
Equivalence: clocks slow down in a gravitational field.
When identical clocks are
compared at different locations in a gravitational field, the
lower clock ticks slightly slower, its frequency $\nu$ being reduced by
\begin{equation}
\frac{\delta \nu}{\nu}=\frac{\Delta {\cal U}}{c^2},
\label{delf}
\end{equation}
where $\Delta {\cal U}$ is the gravitational potential difference
between
the locations of the clocks and $c,$ the speed of light.
The effect has been
experimentally measured using clocks on a tower\cite{tower}, an
aircraft\cite{aircraft} and a rocket\cite{rocket}. More recently
the experiment of Chou {\it et al} ~\cite{chou} measures the
gravitational redshift effect, by comparing two ${\rm
Al^+}$ ion clocks separated in height by just $33{\rm cm}$. The
gravitational redshift (GRS) is at
the foundation of Einstein's General Relativity (GR) and
supports the idea that gravity is encoded in the curvature of spacetime.
There is every reason to measure the gravitational redshift with
as much accuracy as possible.
Indeed, there is a proposal \cite{aces} to further improve the
accuracy
by putting an atomic clock ensemble in space.

In a recent paper M\"uller, Peters and Chu\cite{mpc} (MPC) have
suggested that existing experiments on atom
interferometry~\cite{PhysRevLett.67.181,naturedecade} can be   
reinterpreted as a sensitive test of the gravitational redshift effect.
If this claim is correct, one could achieve high accuracy   
without the trouble and expense of a space mission.  
The claim was based on the Compton
frequency of an atomic mass $m$: one writes $E = mc^{2}=h\nu$ and arrives
at
$\nu_{\rm Compton} = \frac{mc^{2}}{h}$.
The advantage of a Compton frequency clock is that it ticks
at the frequency $\sim10^{25}{\rm Hz}$
which is considerably (about $10^{10}$ times) higher than the
optical frequencies.
As a result $\delta \nu$ in equation (\ref{delf}) is larger for higher
$\nu$ and easier to detect as a fringe shift in interferometry.

However, the claim of MPC is disputed by Wolf {\it et al}
(WBBRSC) \cite{wolf1,wolf2}, who note that the atom interferometer
experiments only constitute a test of the Universality of  Free Fall   
(UFF) and not of the GRS.  Since the gravitational redshift
applies to {\it all} clocks, it is also referred to  
as the Universality of Clock Rates (UCR).
WBBRSC object to MPC's claim
on the grounds that a detailed
analysis \cite{storey} of atom interferometer experiments shows that
the Compton frequency does not appear in the
final answer for the calculated fringe shift.
The analysis presented in \cite{storey} is performed for quadratic
Lagrangians describing the atoms and the propagator is explicitly
calculated. 
WBBRSC also suggest~\cite{wolf2} that the GRS requires
a continuous exchange of signals between the participating
clocks and such exchange would be equivalent to {\it welcherweg}
information which destroys the interference pattern.
They argue that the atom does not deliver a physical
signal at the Compton frequency.
MPC, however, stand by \cite{mpcreply} their claim,  
which is approvingly quoted by the authors of \cite{poli}
and repeated in \cite{hohensee,hohensee2}, which
include some of MPC as co-authors. The matter evidently
is not settled.

Our purpose here is to incisively confront the claim
of MPC by sharpening the objections raised by
WBBRSC. Of the objections raised by WBBRSC,
one of them stands out as being fundamental: that there is nothing
physical about the Compton frequency in this experiment. We will
rest our case entirely on this objection. We will theoretically
examine some conceptual questions raised by this controversy.
We start by critically examining in section II the notion of a ``clock'' 
in 
general 
relativity. In section III, we theoretically analyze an atom 
interferometry experiment and show that it does not test the redshift
at the Compton frequency. In section IV, we propose a 
``clock interferometry'' experiment which does test the redshift, though 
not at the Compton frequency. Finally in the discussion in section V, we 
make a number of comments and note that our proposed experiment is a 
quantum version of the classical clock ``paradox''.

\section{What is a clock?}

Einstein's principle of equivalence implies both the
Universality
of Free Fall (UFF)and the Universality of Clock Rates (UCR). To test
the principle of equivalence, it
is important to test {\it both} these effects independently.
The Universality of Free Fall (UFF) can
be tested by dropping masses, as in Galileo's famous experiment
or by constructing sensitive torsion balances with suspended masses, as
in the E\"otv\"os experiments.
For testing the GRS (or the UCR), it is evident that one {\it needs} to
have clocks not just masses.
What, then, is a clock? A clock is anything which
ticks--delivers a periodic signal. It is usually a dynamical system
which
executes a periodic motion like a pendulum, a planetary orbit, the moons
of
Jupiter or a crystal oscillator. The period defines the ``ticks''of the
clock,
which gives the
least count in time measurement.
Precise clocks have high tick rates. The most precise
clocks in use today are atomic clocks operating at optical frequencies,
ticking at the rate of $10^{15}{\rm Hz}$. These clocks operate in a
quantum
superposition of nondegenerate energy eigenstates
\begin{equation}
|{\psi (t)}> = e^{-i\frac{E_{1}t}{\hbar}} |{\psi_{1}}> +
e^{-i\frac{E_{2}t}{\hbar}} |{\psi_{2}}>
\label{superpose}
\end{equation}
where $|{\psi_{1}}>$ and $|{\psi_{2}}>$ are stationary states
(eigenstates of the energy) and $E_{1}$ and $E_{2}$ the corresponding
distinct energies.
The oscillation frequency of the atomic clock is given by the {\it
difference}  
of the two energies
\begin{equation}
\nu_{clock} = \frac{E_{2} - E_{1}}{h}
\end{equation}
For atoms, typical differences in energy (called spectral terms in 
spectroscopy) are in the ${\rm ev}$ range.
If this frequency is in the optical range, the ticks of the clock are
at $10^{15}{\rm Hz}$. This is an improvement over microwave clocks, which
operate at a lower frequency. What makes a quantum clock tick is
the superposition of {\it at least two} stationary states.
The ticking rate is given by the beat frequency between these states.
Classical clocks like crystal oscillators are in fact superpositions of
many highly excited stationary states and can be described in quantum
mechanics as coherent states.
In contrast, an atom in a stationary state
\begin{equation}   
|{\tilde{\psi}(t)}> = e^{-i\frac{E_{1}t}{\hbar}} |{\psi_{1}}>
\label{stationary}
\end{equation}
is not a clock because it does not execute any periodic motion.    
While the wave function in equation (\ref{stationary}) solving
Schr\"{o}dinger's
equation does appear to have periodic
time dependence, it is important to realise that the wave function is not
directly observable: only bilinears
constructed from the wave function  are.
Thus, bilinears constructed from the wave function in
equation (\ref{stationary})
would be {\it
time independent} while bilinears made from equation (\ref{superpose})
would have
the interference term
\begin{equation}
2{\rm Re} \big[e^{-i\nu t} \psi^*_1(t,r) \psi_2(t,r)\big],   
\end{equation}
which is a measurable quantity reflecting the oscillating charge density
of
the electronic motion within the atom. In fact, this oscillation
leads to an oscillating dipole moment for the atom, which couples to
radiation during atomic transitions. 

\begin{figure}[fig1]
\centering \vspace*{3mm} \hspace*{1mm}  
\includegraphics[width =0.6\columnwidth]{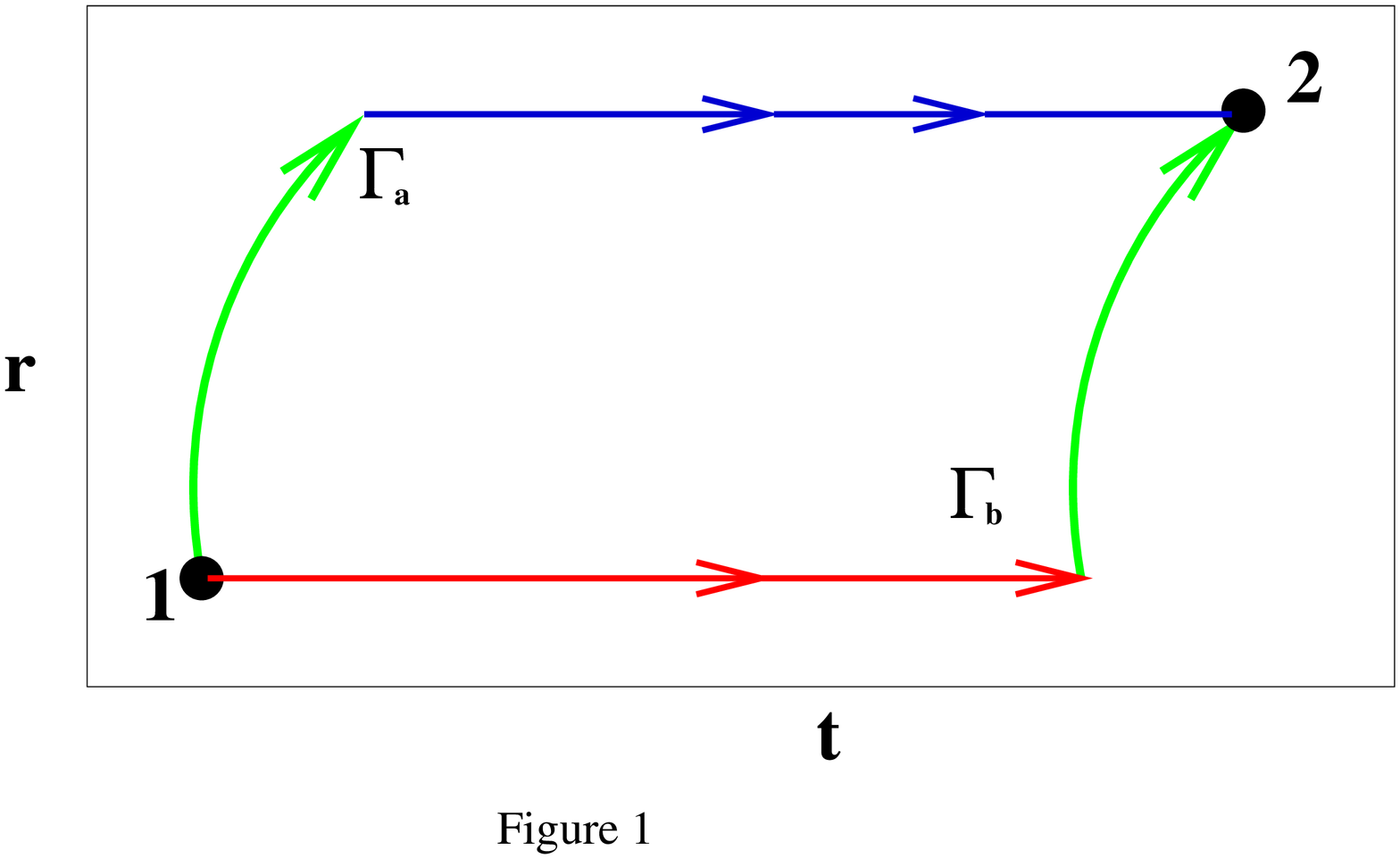} 
\vspace*{-1mm}
\caption{Schematic diagram of the atom interferometry experiment 
(colour online):
${\bf 1}$ and ${\bf 2}$ are respectively the source and detector of   
atoms and $\Gamma_a$ and $\Gamma_b$, two paths in spacetime connecting
the source and the detector. Time
$t$ is
plotted horizontally and $r$ the radial coordinate is     
plotted vertically.
The horizontal lower (red online) and upper (blue online) lines
represent the locations of the lower and higher
atom traps in which the atom resides. The two rising (green online) 
lines in Figure 1 are the
spacetime paths by which the atoms are raised to the higher trap.
Atoms traversing the path $\Gamma_b$ spend more time deeper in the
gravitational potential than those traversing
$\Gamma_a$. There is a relative phase shift
${\Delta \phi}=\frac{m_g g z T}{\hbar}$, with $z$ the height difference
and $T$ the coordinate time of residence of both atoms in their traps.
}
\label{fig1}
\end{figure}

MPC suggest that an atom is a clock at the
Compton frequency.
We contradict this suggestion and show by
general arguments
that the Compton frequency is not a gauge invariant observable in the
atom interferometry experiment. Atoms in stationary states are {\it 
masses}
and do not {\it tick}. They cannot be viewed as clocks ticking at the 
Compton
rate.

\section{Quantum Interference of Atoms:}

Consider the interference between atoms emitted at $1$ and received
at $2$(Figure $1$). If an atom is initially in an atom trap at 1 
(in red online) and
is moved with a probability amplitude $1/\sqrt{2}$ to a higher trap (in
blue online), one
can observe interference between the two possible histories $\Gamma_a$
and $\Gamma_b$. Our analysis below 
shows that the effects of the Compton
frequency  can be eliminated in all observables.
{\it Thus an atom in a stationary state is not a clock ticking at the
Compton frequency.}
It takes two energy states to beat, just as it takes two hands to clap.


The experiment may be described either by the
Schr\"{o}dinger
equation or the equivalent Feynman path integral.
The phase picked up by the atom in following a path $\Gamma$ is
$S_\Gamma/\hbar$, where
\begin{equation}
S_{\Gamma} = \int^{}_{\Gamma}\big[(-mc)ds - V(t,r)dt\big]
\end{equation}
where $ds$ is the proper interval measured along $\Gamma$ and $V(t,r)$
is the effect of non gravitational potentials which may be used to
manipulate the atom. The gravitational field of the earth can be     
described by the Schwarzschild metric
\begin{equation}
ds^{2} = (1-\frac{2GM}{c^2r}) c^2dt^{2} - (1-\frac{2GM}{c^2r})^{-1}
dr^{2} - r^{2}d\Omega^{2},
\end{equation}
and to the required accuracy, the Lagrangian describing the motion of
the atom is
\begin{equation}
L = -mc^{2} + \frac{1}{2}m \big(\frac{dr}{dt}\big)^{2} +
\frac{GMm}{r} -V(t,r),
\end{equation}
which leads to the Hamiltonian
\begin{equation}
H= \frac{p^{2}}{2m} + V(t,r) - \frac{GMm}{r} + mc^{2}.
\label{hamiltonian}
\end{equation}  

(Strictly, since we are testing the theory, we ought to go beyond 
GR and allow for the possibility that the masses appearing in the potential 
and kinetic terms differ. We gloss over this 
point since our interest is only in the effect of the constant term.)

The energy corresponding to the ``Compton frequency'' is present in the 
Hamiltonian (\ref{hamiltonian}) as an additive
constant $mc^{2}$. This term cancels out in all energy differences
(or spectral terms) and is therefore unobservable.
The unobservability of the constant is driven home by noticing that the
problem admits a ``gauge symmetry''.
If $\psi(t,r)$ is a solution
then
\begin{equation}
\psi^{'} (t,r)=U(t,r) \psi (t,r),
\label{gauge}
\end{equation} 
where $U(t,r) =\exp i \chi (t,r)$ 
is a solution of the Schr$\ddot{\rm o}$dinger  equation with
Hamiltonian     
\begin{equation}
H^{'} = UH U^{-1}  + i \hbar  \dot{U}U^{-1}.\\
\label{htransform}
\end{equation}

The choice $\chi(t,r)=\frac{mc^2t}{\hbar}$
results in a new Hamiltonian
\begin{equation}
H^{'} = \frac{p^{2}}{2m} - \frac{GMm}{r} + V(t,r),
\end{equation}
in which the Compton frequency disappears.
Since one only measures bilinears in the wave function, the gauge argument
shows quite generally that the effects of the Compton
frequency  can be eliminated in all observables.

The argument can easily be translated to the Feynman path integral
formalism. 
Under a gauge transformation the Lagrangian changes by a total time
derivative
\begin{equation}  
L^{'} = L+\frac{d\chi}{dt}
\end{equation}
and as a result we find that the propagator
\begin{equation}
K(t_1,r_1;t_2,r_2)=\Sigma_{\Gamma} \exp{iS_\Gamma}
\label{Feynman}
\end{equation}
(expressed as a Feynman path integral over all spacetime paths $\Gamma$
which go from $(t_1,r_1)$ to $(t_2,r_2)$)
transforms as
\begin{equation}
K'(t_1,r_1;t_2,r_2)=\exp{i\chi(t_1,r_1)}K(t_1,r_1;t_2,r_2)exp{-i\chi(t_2,r_2)}
\label{feynman}
\end{equation}
All physical results are unchanged. The freedom to add a constant  
to the Hamiltonian is present for particles in external fields,
gravitational or otherwise. This freedom is lost only when one
considers the gravitational field of the particle itself, in this
case the atom. This gravitational field is clearly negligible in
the present context.
We conclude that the rest mass of an atom in this experiment does not
deliver a physical signal at the Compton frequency and therefore 
an atom is {\it not} a clock ticking at the Compton frequency.

In the gedanken experiment shown
in Figure 1, the phase difference
of the atoms arriving at the detector is easily calculated. 
One observes interference between two histories $\Gamma_a$
and $\Gamma_b$, each of which has equal amplitude $\frac{1}{\sqrt{2}}$.
In $\Gamma_a$, the atom is moved to a higher trap and
spends coordinate time $T$
in the higher trap, while in $\Gamma_b$, the atom stays in the lower trap
for a time $T$ and is then lifted to the higher trap by an external force
(supplied by lasers). We can suppose that the non gravitational
potential has the same value $0$ in both traps and that the phases
picked up in the rising (green online) sectors cancel exactly. 
Since the atoms are stationary in the traps,
there is no effect of
motion and for a residence time $T$, the phase difference is  
$mT/\hbar$ times the
difference of the
gravitational potential $\cal U$ between the two traps separated in
height by $z$. The final
answer is
\begin{equation}
\delta \phi=\frac{m\Delta {\cal U}T}{\hbar}\approx\frac{mgzT}{\hbar}
\label{final}
\end{equation}
which is in {\it complete agreement with MPC}. It is only the 
interpretation of this result as a detection of the redshift effect 
at the Compton frequency that we dispute.
We interpret Equation (\ref{final}) as the phase shift due to 
fall under gravity $g$, not as a redshift. This is because 
the atoms are in stationary states and therefore are not clocks.

Our result (\ref{final}) and that of MPC are apparently at variance with 
that of \cite{wolf1,wolf2}. The difference is easily understood. 
Refs.\cite{wolf1,wolf2} compute the quantum propagator explicitly,
which is only possible in the quadratic approximation. 
In order to better compare our results with those of 
\cite{wolf1,wolf2,storey} 
we assume that $V$ is time independent and expand $V(r)+\frac{GM}{r}$ 
in a Taylor series around  its minimum at $r_0$
to quadratic order in $(r-r_0)$. We find the Lagrangian  
\begin{equation}
L = -mc^{2} + \frac{1}{2}m \big(\frac{dr}{dt}\big)^{2} -
\frac{k}{2}(r-r_0)^2-V(r_0),
\end{equation}
where $k$ is an effective spring constant. This Lagrangian describes 
the simple harmonic oscillator.

In order to cause interference we must consider two classical 
histories that start and end at the same spacetime point. 
For example $r_1(t) = r_0+A_1 \sin\omega t$
and $r_2(t) = r_0+A_2 \sin\omega t$ which intersect at $t=0$ and 
$t=\frac{2\pi}{\omega}$.
Such pairs are conjugate points and as is well known, the total 
phase acquired by an oscillator is independent of the amplitude. The 
phase difference is therefore zero, in agreement with \cite{storey}. 
There is no conflict, however, between the phase shift calculations 
of MPC and \cite{wolf1,wolf2,storey}. These computations apply to {\it 
different} situations and they are {\it both} correct. The phase
shift (\ref{final}) is calculated semiclassically for a 
non-harmonic potential.

Our objection to MPC is {\it not} that their computed phase shift
is incorrect but that their interfering atoms are in stationary states 
and therefore not clocks.

\section{Quantum Interference of Clocks:}
Can one use atom interferometry to test UCR? The answer is yes:
we need to observe the interference of atomic {\it clocks} rather than
atomic {\it masses}. To do this, we must have a source of atoms in
a coherent superposition of different energy states. The phase
picked up by an atom is $\exp{\frac{-i\epsilon\tau}{\hbar}}$,
where $\tau$ is the proper time measured along its worldline
and $\epsilon$ its proper energy. 

Consider an atom in an initial state
$1/\sqrt{2}(|{1}>+|{2}>)$, a superposition of two states with proper
energies $\epsilon_1$ and $\epsilon_2$. A beam splitter at the source
causes the atom to follow paths $\Gamma_a$ and $\Gamma_b$
with equal amplitude. The proper times for the rising (green online), 
upper horizontal (blue online) and
lower horizontal (red online) sections are respectively $\tau$, $\tau_a$ 
and $\tau_b$ 
(see Figure 1).
On arrival at the detector, the state
of the atom is given by a sum of two amplitudes
\begin{equation}
|{\psi_a}>=1/\sqrt{2}(\exp\frac{-i\epsilon_1(\tau+\tau_a)}{2\hbar}|{1}>
+\exp\frac{-i\epsilon_2(\tau+\tau_a)}{2\hbar}|{2}>)
\label{supera}
\end{equation}
and
\begin{equation}
|{\psi_b}>=1/\sqrt{2}(\exp\frac{-i\epsilon_1(\tau+\tau_b)}{2\hbar}|{1}>
+\exp\frac{-i\epsilon_2(\tau+\tau_b)}{2\hbar}|{2}>)
\label{superb}
\end{equation}
The interference between the two alternatives $\Gamma_a$ and $\Gamma_b$
gives a term $2 {\rm Re}<{\psi_a}|{\psi_b}>$ which
is easily computed
to be given by 
\begin{equation}
\big[\cos{{m\Delta {\cal U}T}}/{\hbar}\big]\big[\cos{{\Delta\epsilon}/{\Delta{\cal U}T}{2c^2 \hbar}}\big]
\label{interference}
\end{equation}
where $m=\frac{\epsilon_1+\epsilon_2}{2c^2}$ and
$\Delta\epsilon=\epsilon_1-\epsilon_2$.
The first term in square brackets is the old term in Equation 
[\ref{final}].
However, the phase of the second term is a measurement of the 
redshift effect.

For an atom in a superposition
of states with proper energies $\epsilon_1$ and $\epsilon_2$,
the expected
interference term at the detector in Figure 1 is given by 
Equation (\ref{interference}).
where $m=\frac{\epsilon_1+\epsilon_2}{2c^2}$ and
$\Delta\epsilon=\epsilon_1-\epsilon_2$.
The first factor measures UFF and couples to the mass of the atoms,
while the second measures UCR and couples to the {\it energy difference}
between the superposed states. Experiments of this kind can be  
described as {\it clock interferometry} and have not been done,
to the best of our knowledge.
They can be thought of as a quantum version of the clock ``paradox'',
in which one uses a {\it single clock} in a quantum superposition,
instead of the {\it two}
clocks compared in the classical clock ``paradox'', which is
experimentally realised in Reference \cite{chou}. In the quantum 
version, a
{\it single clock} traverses both alternative world
lines and interferes with itself. Such experiments constitute a good
example of
the use of the internal degrees of freedom of an atom in
interferometry\cite{storey}. They would lead to new tests of
UCR at optical frequencies.

\section{Discussion}

It is known \cite{nortvedt,bondi} that if one assumes UFF and energy
conservation, it follows that UCR must also hold.
This conclusion can be arrived at by considering
systems which can exist in various energy states and make transitions
between them emitting quanta\cite{nortvedt,bondi}.
Indeed, one can construct perpetual motion machines of the first kind
if UFF holds but UCR does not hold. One striking realisation of such  
a machine is Bondi's system of buckets and
pulleys\cite{bondi}, using excited atoms to perpetually outweigh
identical atoms in
their
ground state.
However, more quantitatively, a test of UFF to a certain level of 
accuracy
leads to an accuracy of UCR which is reduced
by a factor $\frac{\Delta E}{E}$ \cite{nortvedt,bondi,will}, where
$E$ is the
absolute energy and $\Delta E$ is the energy
{\it difference} between the states of interest. 
In atomic systems, the energy separations are much
smaller than the rest mass,
$\frac{\Delta E}{E} \sim 10^{-10}$ which is why the present
accuracy of UCR tests are  considerably  lower than those of UFF.

Apart from atomic clocks, which are held together by electromagnetic
interactions, one can also consider clocks held together by other 
forces. Gravitational clocks (like 
the moons of Jupiter) are held together by gravity and their
binding energies are extremely small compared to the mass of the clock.
The clock consists of the entire system,
Jupiter plus its moons, and the clock rate depends on the solar 
gravitational potential. 
In contrast nuclear binding energies 
are often relatively large, 
around $1\%$ of the rest mass and such systems could be explored to
improve the accuracy of UCR tests. 
This point has been noted in Reference \cite{hohensee2} in their recent 
preprint.

We have defined a clock as something which ``ticks'' periodically.
One can also use clocks based on decay rates\cite{nortvedt} or 
transition rates between two energy states.  Our arguments also apply to
such decay clocks, since all we need is that the system must be in a 
superposition of at least two states.

We have been using the word ``atom'' to mean an atom in a stationary 
state and ``atomic clock'' to describe atoms which are in a 
superposition of at least two states. 
Needless to say, the discussion
applies also to ions, which may be easier to manipulate experimentally. 
Our discussion here is at the level of ``gedanken experiments'' and
the translation to a laboratory experiment may involve some changes.

Our final result for the expected phase shift {\it is} in agreement with
Reference \cite{mpc}. Our disagreement is more subtle: we do not agree
with Reference \cite{mpc} that this experiment is a measurement
of the redshift. It only
constitutes
a measurement of the phase shift due to the gravitational acceleration
$g$. Thus the atom interferometry experiment only measures the UFF and 
not UCR. For the latter one needs to have genuine clocks, not just
masses {\it interpreted} as clocks. If one could use atomic masses to
generate a periodic signal, this would lead to unprecedented accuracy in 
time keeping. Present day atomic clocks work at optical frequencies and  
lose no more than a second in the age of the Universe. A Compton
frequency clock would lose no more than a nanosecond
in the age of the Universe.
It seems clear that present technology is far from achieving such
precision in time keeping.

Can one construct clocks at the Compton frequency? The answer, in
principle, is yes. What one has to do is to superpose states which
differ in the number of atoms {\it i.e} generate a quantum state
which is not in a number eigenstate. Such states exist in
a quantum field theory. An example of such a state is a single mode   
coherent state (with $Z$ a complex number),
$|{Z}>=\exp{-|Z|^2}\Sigma_{0}^{\infty} {Z^n}/{\sqrt{n!}}|{n}>$
which superposes different numbers of particles.
A laser beam is an example of a coherent photon field which
can be regarded as a clock at optical frequencies. Needless to say,
experiments using superpositions of atom numbers in interferometry
have not so far been performed.

To summarize, Einstein's Equivalence Principle implies Universality
of Free Fall (UFF) and Universality of Clock Rates (UCR). It is 
important to test {\it both} of these. Tests of UFF
entail the use of masses, whereas one {\it needs} clocks to check
UCR. In this article we explicitly show that an atom is not a clock  
ticking at the Compton frequency and therefore one cannot achieve an
advantage of ten orders of magnitude in precision compared to
existing tests of UCR.

In the popular relativity literature, considerable attention has been 
devoted to the classical twin ``paradox''. 
Two twins are separated at birth and follow different world lines.
(In popular accounts, one stays home.) When they meet after some 
years, one has aged relative to the other. The two twins
evidently carry biological clocks, 
which are synchronised at birth
and compared when they meet. (There is no need for the twins
to exhange signals between these events.) In an actual experiment
Chou et al \cite{chou} have realised this effect using two ${\rm Al}^+$
ions to play the role of the two twins. Like the twins,
the ions are clocks and clocks ``age'' differently on different
world lines. (It is of course necessary that the ions are in a 
superposition of at least two stationary states so that they tick!)
It is possible to come up with a quantum version
of this effect: we do not need {\it two} ``twins''. Starting 
with a single ion source of ion clocks (these have to be in a 
superposition of internal states to qualify as clocks)
we perform a split beam experiment so that the single ion has
equal amplitude to traverse the two arms of an 
interferometer. On recombining the beams, the interference 
between the arms will reveal the presence of differential aging.

Our proposed experiment in Figure 1 to measure GRS is exactly of this 
kind. Atomic clocks separated by a beam splitter into two 
beams, one of which lies
deeper in a gravitational field than the other. When 
the beams are recombined their interference will reveal the 
presence (\ref{interference})
of a different number of ticks in the two arms.
Of course it is necessary that the atoms are genuine clocks,
{\it i.e} they must be in a superposition of internal energy states. 
GRS measurements can thus be performed in 
atom interferometry by causing quantum interference between 
atomic {\it clocks}. We hope to interest the atom interferometric
community in developing a realisation of this gedanken experiment.

\vskip 1cm \noindent{\bf Acknowledgments :}
It is a pleasure to thank Luc Blanchet, whose colloquium on this
subject excited our interest, for his critical comments
and corrections. We gratefully acknowledge comments by Wei-Tou Ni.
We have benefitted from criticism from Holger M\"uller,
who holds a different view.

 


\end{document}